%Paper: cond-mat/9408098
%From: Nobuki Maeda <maeda@phys.hokudai.ac.jp>
%Date: Wed, 31 Aug 94 19:32:23 JST

%%%%%%%%%%%%%%%%%%%%%%%%%%%%%%%%%%%%%%%%%%%%%%%%%%%%%%%%%%%%%%%%%%%%%%
% Table is written by LaTeX
%%%%%%%%%%%%%%%%%%%%%%%%%%%%%%%%%%%%%%%%%%%%%%%%%%%%%%%%%%%%%%%%%%%%%%
\expandafter\ifx\csname phyzzx\endcsname\relax
 \message{It is better to use PHYZZX format than to
          \string\input\space PHYZZX}\else
 \wlog{PHYZZX macros are already loaded and are not
          \string\input\space again}%
 \endinput \fi
\catcode`\@=11 % This allows us to modify PLAIN macros.
\let\rel@x=\relax
\let\n@expand=\relax
\def\pr@tect{\let\n@expand=\noexpand}
\let\protect=\pr@tect
\let\gl@bal=\global

\newfam\cpfam
\newdimen\b@gheight             \b@gheight=12pt
\newcount\f@ntkey               \f@ntkey=0
\def\f@m{\afterassignment\samef@nt\f@ntkey=}
\def\samef@nt{\fam=\f@ntkey \the\textfont\f@ntkey\rel@x}
\def\setstr@t{\setbox\strutbox=\hbox{\vrule height 0.85\b@gheight
                                depth 0.35\b@gheight width\z@ }}
\input phyzzx.fonts
% Actual font definitions are kept in a separate file
% to facilitate font substitution.
%
\def\rm{\n@expand\f@m0 }
\def\mit{\n@expand\f@m1 }         
\def\cal{\n@expand\f@m2 }
\def\it{\n@expand\f@m\itfam}
\def\sl{\n@expand\f@m\slfam}
\def\bf{\n@expand\f@m\bffam}
\def\tt{\n@expand\f@m\ttfam}
\def\caps{\n@expand\f@m\cpfam}    
\def\em@{\rel@x\ifnum\f@ntkey=0 \it \else
        \ifnum\f@ntkey=\bffam \it \else \rm \fi \fi }
\def\em{\n@expand\em@}
\def\fourteenpoint{\fourteenf@nts \samef@nt \b@gheight=14pt \setstr@t }
\def\twelvepoint{\twelvef@nts \samef@nt \b@gheight=12pt \setstr@t }
\def\tenpoint{\tenf@nts \samef@nt \b@gheight=10pt \setstr@t }
\normalbaselineskip = 19.2pt plus 0.2pt minus 0.1pt %xxx not 20pt
\normallineskip = 1.5pt plus 0.1pt minus 0.1pt
\normallineskiplimit = 1.5pt
\newskip\normaldisplayskip
\normaldisplayskip = 14.4pt plus 3.6pt minus 10.0pt %xxx not 20 +5 -10
\newskip\normaldispshortskip
\normaldispshortskip = 6pt plus 5pt
\newskip\normalparskip
\normalparskip = 6pt plus 2pt minus 1pt
\newskip\skipregister
\skipregister = 5pt plus 2pt minus 1.5pt
\newif\ifsingl@
\newif\ifdoubl@
\newif\iftwelv@  \twelv@true
\def\singlespace{\singl@true\doubl@false\spaces@t}
\def\doublespace{\singl@false\doubl@true\spaces@t}
\def\normalspace{\singl@false\doubl@false\spaces@t}
\def\Tenpoint{\tenpoint\twelv@false\spaces@t}
\def\Twelvepoint{\twelvepoint\twelv@true\spaces@t}
\def\spaces@t{\rel@x
      \iftwelv@ \ifsingl@\subspaces@t3:4;\else\subspaces@t1:1;\fi
       \else \ifsingl@\subspaces@t3:5;\else\subspaces@t4:5;\fi \fi
      \ifdoubl@ \multiply\baselineskip by 5
         \divide\baselineskip by 4 \fi }
\def\subspaces@t#1:#2;{
      \baselineskip = \normalbaselineskip
      \multiply\baselineskip by #1 \divide\baselineskip by #2
      \lineskip = \normallineskip
      \multiply\lineskip by #1 \divide\lineskip by #2
      \lineskiplimit = \normallineskiplimit
      \multiply\lineskiplimit by #1 \divide\lineskiplimit by #2
      \parskip = \normalparskip
      \multiply\parskip by #1 \divide\parskip by #2
      \abovedisplayskip = \normaldisplayskip
      \multiply\abovedisplayskip by #1 \divide\abovedisplayskip by #2
      \belowdisplayskip = \abovedisplayskip
      \abovedisplayshortskip = \normaldispshortskip
      \multiply\abovedisplayshortskip by #1
        \divide\abovedisplayshortskip by #2
      \belowdisplayshortskip = \abovedisplayshortskip
      \advance\belowdisplayshortskip by \belowdisplayskip
      \divide\belowdisplayshortskip by 2
      \smallskipamount = \skipregister
      \multiply\smallskipamount by #1 \divide\smallskipamount by #2
      \medskipamount = \smallskipamount \multiply\medskipamount by 2
      \bigskipamount = \smallskipamount \multiply\bigskipamount by 4 }
\def\normalbaselines{ \baselineskip=\normalbaselineskip
   \lineskip=\normallineskip \lineskiplimit=\normallineskip
   \iftwelv@\else \multiply\baselineskip by 4 \divide\baselineskip by 5
     \multiply\lineskiplimit by 4 \divide\lineskiplimit by 5
     \multiply\lineskip by 4 \divide\lineskip by 5 \fi }
\Twelvepoint  % That's the default
\interlinepenalty=50
\interfootnotelinepenalty=5000
\predisplaypenalty=9000
\postdisplaypenalty=500
\hfuzz=1pt
\vfuzz=0.2pt
\newdimen\HOFFSET  \HOFFSET=0pt
\newdimen\VOFFSET  \VOFFSET=0pt
\newdimen\HSWING   \HSWING=0pt
\dimen\footins=8in

\newskip\pagebottomfiller
\pagebottomfiller=\z@ plus \z@ minus \z@
\def\pagecontents{
   \ifvoid\topins\else\unvbox\topins\vskip\skip\topins\fi
   \dimen@ = \dp255 \unvbox255
   \vskip\pagebottomfiller
   \ifvoid\footins\else\vskip\skip\footins\footrule\unvbox\footins\fi
   \ifr@ggedbottom \kern-\dimen@ \vfil \fi }
\def\makeheadline{\vbox to 0pt{ \skip@=\topskip
      \advance\skip@ by -12pt \advance\skip@ by -2\normalbaselineskip
      \vskip\skip@ \line{\vbox to 12pt{}\the\headline} \vss
      }\nointerlineskip}
\def\makefootline{\baselineskip = 1.5\normalbaselineskip
                 \line{\the\footline}}
\newif\iffrontpage
\newif\ifp@genum
\def\nopagenumbers{\p@genumfalse}
\def\pagenumbers{\p@genumtrue}
\pagenumbers
\newtoks\paperheadline
\newtoks\paperfootline
\newtoks\letterheadline
\newtoks\letterfootline
\newtoks\letterinfo
\newtoks\date
\paperheadline={\hfil}
\paperfootline={\hss\iffrontpage\else\ifp@genum\tenrm\folio\hss\fi\fi}
\letterheadline{\iffrontpage \hfil \else
    \rm \ifp@genum page~~\folio\fi \hfil\the\date \fi}
\letterfootline={\iffrontpage\the\letterinfo\else\hfil\fi}
\letterinfo={\hfil}
\def\monthname{\rel@x\ifcase\month 0/\or January\or February\or
   March\or April\or May\or June\or July\or August\or September\or
   October\or November\or December\else\number\month/\fi}
\def\today{\monthname~\number\day, \number\year}
\date={\today}
\headline=\paperheadline % The default is
\footline=\paperfootline % \papers
\countdef\pageno=1      \countdef\pagen@=0
\countdef\pagenumber=1  \pagenumber=1
\def\advancepageno{\gl@bal\advance\pagen@ by 1
   \ifnum\pagenumber<0 \gl@bal\advance\pagenumber by -1
    \else\gl@bal\advance\pagenumber by 1 \fi
    \gl@bal\frontpagefalse  \swing@ }
\def\folio{\ifnum\pagenumber<0 \romannumeral-\pagenumber
           \else \number\pagenumber \fi }
\def\swing@{\ifodd\pagenumber \gl@bal\advance\hoffset by -\HSWING
             \else \gl@bal\advance\hoffset by \HSWING \fi }
\def\footrule{\dimen@=\prevdepth\nointerlineskip
   \vbox to 0pt{\vskip -0.25\baselineskip \hrule width 0.35\hsize \vss}
   \prevdepth=\dimen@ }
\let\footnotespecial=\rel@x
\newdimen\footindent
\footindent=24pt
\def\Textindent#1{\noindent\llap{#1\enspace}\ignorespaces}
\def\Vfootnote#1{\insert\footins\bgroup
   \interlinepenalty=\interfootnotelinepenalty \floatingpenalty=20000
   \singl@true\doubl@false\Tenpoint
   \splittopskip=\ht\strutbox \boxmaxdepth=\dp\strutbox
   \leftskip=\footindent \rightskip=\z@skip
   \parindent=0.5\footindent \parfillskip=0pt plus 1fil
   \spaceskip=\z@skip \xspaceskip=\z@skip \footnotespecial
   \Textindent{#1}\footstrut\futurelet\next\fo@t}

\def\vfootnote#1{\Vfootnote{${#1}$}}
\def\footnote#1{\attach{#1}\vfootnote{#1}}

\let\footsymbol=\star
\newcount\lastf@@t           \lastf@@t=-1
\newcount\footsymbolcount    \footsymbolcount=0
\newif\ifPhysRev
\def\bumpfootsymbolcount{\rel@x
   \iffrontpage \bumpfootsymbolpos \else \advance\lastf@@t by 1
     \ifPhysRev \bumpfootsymbolneg \else \bumpfootsymbolpos \fi \fi
   \gl@bal\lastf@@t=\pagen@ }
\def\bumpfootsymbolpos{\ifnum\footsymbolcount <0
                            \gl@bal\footsymbolcount =0 \fi
    \ifnum\lastf@@t<\pagen@ \gl@bal\footsymbolcount=0
     \else \gl@bal\advance\footsymbolcount by 1 \fi }
\def\bumpfootsymbolneg{\ifnum\footsymbolcount >0
             \gl@bal\footsymbolcount =0 \fi
         \gl@bal\advance\footsymbolcount by -1 }
\def\fd@f#1 {\xdef\footsymbol{\mathchar"#1 }}
\def\generatefootsymbol{\ifcase\footsymbolcount \fd@f 13F \or \fd@f 279
        \or \fd@f 27A \or \fd@f 278 \or \fd@f 27B \else
        \ifnum\footsymbolcount <0 \fd@f{023 \number-\footsymbolcount }
         \else \fd@f 203 {\loop \ifnum\footsymbolcount >5
                \fd@f{203 \footsymbol } \advance\footsymbolcount by -1
                \repeat }\fi \fi }

\def\nonfrenchspacing{\sfcode`\.=3001 \sfcode`\!=3000 \sfcode`\?=3000
        \sfcode`\:=2000 \sfcode`\;=1500 \sfcode`\,=1251 }
\nonfrenchspacing
\newdimen\d@twidth
{\setbox0=\hbox{s.} \gl@bal\d@twidth=\wd0 \setbox0=\hbox{s}
        \gl@bal\advance\d@twidth by -\wd0 }
\def\removehglue{\loop \unskip \ifdim\lastskip >\z@ \repeat }
\def\roll@ver#1{\removehglue \nobreak \count255 =\spacefactor \dimen@=\z@
        \ifnum\count255 =3001 \dimen@=\d@twidth \fi
        \ifnum\count255 =1251 \dimen@=\d@twidth \fi
    \iftwelv@ \kern-\dimen@ \else \kern-0.83\dimen@ \fi
   #1\spacefactor=\count255 }
\def\step@ver#1{\rel@x \ifmmode #1\else \ifhmode
        \roll@ver{${}#1$}\else {\setbox0=\hbox{${}#1$}}\fi\fi }
\def\attach#1{\step@ver{\strut^{\mkern 2mu #1} }}

\newcount\chapternumber      \chapternumber=0
\newcount\sectionnumber      \sectionnumber=0
\newcount\equanumber         \equanumber=0
\let\chapterlabel=\rel@x
\let\sectionlabel=\rel@x
\newtoks\chapterstyle        \chapterstyle={\Number}
\newtoks\sectionstyle        \sectionstyle={\chapterlabel.\Number}
\newskip\chapterskip         \chapterskip=\bigskipamount
\newskip\sectionskip         \sectionskip=\medskipamount
\newskip\headskip            \headskip=8pt plus 3pt minus 3pt
\newdimen\chapterminspace    \chapterminspace=15pc
\newdimen\sectionminspace    \sectionminspace=10pc
\newdimen\referenceminspace  \referenceminspace=20pc
\def\chapterreset{\gl@bal\advance\chapternumber by 1
   \ifnum\equanumber<0 \else\gl@bal\equanumber=0\fi
   \sectionnumber=0 \let\sectionlabel=\rel@x
   {\pr@tect\xdef\chapterlabel{\the\chapterstyle{\the\chapternumber}}}}
\def\alphabetic#1{\count255='140 \advance\count255 by #1\char\count255}
\def\Alphabetic#1{\count255='100 \advance\count255 by #1\char\count255}
\def\Roman#1{\uppercase\expandafter{\romannumeral #1}}
\def\roman#1{\romannumeral #1}
\def\Number#1{\number #1}
\def\BLANC#1{}
\def\titleparagraphs{\interlinepenalty=9999
     \leftskip=0.03\hsize plus 0.22\hsize minus 0.03\hsize
     \rightskip=\leftskip \parfillskip=0pt
     \hyphenpenalty=9000 \exhyphenpenalty=9000
     \tolerance=9999 \pretolerance=9000
     \spaceskip=0.333em \xspaceskip=0.5em }
\def\titlestyle#1{\par\begingroup \titleparagraphs
     \iftwelv@\fourteenpoint\else\twelvepoint\fi
   \noindent #1\par\endgroup }
\def\spacecheck#1{\dimen@=\pagegoal\advance\dimen@ by -\pagetotal
   \ifdim\dimen@<#1 \ifdim\dimen@>0pt \vfil\break \fi\fi}
\def\chapter#1{\par \penalty-300 \vskip\chapterskip
   \spacecheck\chapterminspace
   \chapterreset \titlestyle{\chapterlabel.~#1}
   \nobreak\vskip\headskip \penalty 30000
   {\pr@tect\wlog{\string\chapter\space \chapterlabel}} }

\def\section#1{\par \ifnum\the\lastpenalty=30000\else
   \penalty-200\vskip\sectionskip \spacecheck\sectionminspace\fi
   \gl@bal\advance\sectionnumber by 1
   {\pr@tect
   \xdef\sectionlabel{\the\sectionstyle\the\sectionnumber}
   \wlog{\string\section\space \sectionlabel}}
   \noindent {\caps\enspace\sectionlabel.~~#1}\par
   \nobreak\vskip\headskip \penalty 30000 }
\def\subsection#1{\par
   \ifnum\the\lastpenalty=30000\else \penalty-100\smallskip \fi
   \noindent\undertext{#1}\enspace \vadjust{\penalty5000}}

\def\undertext#1{\vtop{\hbox{#1}\kern 1pt \hrule}}
\def\APPENDIX#1#2{\par\penalty-300\vskip\chapterskip
   \spacecheck\chapterminspace \chapterreset \xdef\chapterlabel{#1}
   \titlestyle{APPENDIX #2} \nobreak\vskip\headskip \penalty 30000
   \wlog{\string\Appendix~\chapterlabel} }
\def\Appendix#1{\APPENDIX{#1}{#1}}
\def\appendix{\APPENDIX{A}{}}
\def\unnumberedchapters{\let\makechapterlabel=\rel@x
      \let\chapterlabel=\rel@x  \sectionstyle={\BLANC}
      \let\sectionlabel=\rel@x \sequentialequations }

\def\eqname#1{\rel@x {\pr@tect
  \ifnum\equanumber<0 \xdef#1{{\rm(\number-\equanumber)}}%
     \gl@bal\advance\equanumber by -1
  \else \gl@bal\advance\equanumber by 1
     \ifx\chapterlabel\rel@x \def\d@t{}\else \def\d@t{.}\fi
    \xdef#1{{\rm(\chapterlabel\d@t\number\equanumber)}}\fi #1}}
\def\eq{\eqname\?}

\def\eqinsert#1{\noalign{\dimen@=\prevdepth \nointerlineskip
   \setbox0=\hbox to\displaywidth{\hfil #1}
   \vbox to 0pt{\kern 0.5\baselineskip\hbox{$\!\box0\!$}\vss}
   \prevdepth=\dimen@}}

\def\GENITEM#1;#2{\par \hangafter=0 \hangindent=#1
    \Textindent{$ #2 $}\ignorespaces}
\outer\def\newitem#1=#2;{\gdef#1{\GENITEM #2;}}

\newdimen\itemsize                \itemsize=30pt
\newitem\item=1\itemsize;
\newitem\sitem=1.75\itemsize;     
\newitem\ssitem=2.5\itemsize;     
\outer\def\newlist#1=#2&#3&#4;{\toks0={#2}\toks1={#3}%
   \count255=\escapechar \escapechar=-1
   \alloc@0\list\countdef\insc@unt\listcount     \listcount=0
   \edef#1{\par
      \countdef\listcount=\the\allocationnumber
      \advance\listcount by 1
      \hangafter=0 \hangindent=#4
      \Textindent{\the\toks0{\listcount}\the\toks1}}
   \expandafter\expandafter\expandafter
    \edef\c@t#1{begin}{\par
      \countdef\listcount=\the\allocationnumber \listcount=1
      \hangafter=0 \hangindent=#4
      \Textindent{\the\toks0{\listcount}\the\toks1}}
   \expandafter\expandafter\expandafter
    \edef\c@t#1{con}{\par \hangafter=0 \hangindent=#4 \noindent}
   \escapechar=\count255}
\def\c@t#1#2{\csname\string#1#2\endcsname}
\newlist\point=\Number&.&1.0\itemsize;
\newlist\subpoint=(\alphabetic&)&1.75\itemsize;
\newlist\subsubpoint=(\roman&)&2.5\itemsize;

\newcount\referencecount     \referencecount=0
\newcount\lastrefsbegincount \lastrefsbegincount=0
\newif\ifreferenceopen       \newwrite\referencewrite
\newdimen\refindent          \refindent=30pt
\def\normalrefmark#1{\attach{\scriptscriptstyle [ #1 ] }}
\let\PRrefmark=\attach
\def\NPrefmark#1{\step@ver{{\;[#1]}}}
\def\refmark#1{\rel@x\ifPhysRev\PRrefmark{#1}\else\normalrefmark{#1}\fi}
\def\refend@{\refmark{\number\referencecount}}
\def\refend{\refend@{}\space }
\def\refsend{\refmark{\count255=\referencecount
   \advance\count255 by-\lastrefsbegincount
   \ifcase\count255 \number\referencecount
   \or \number\lastrefsbegincount,\number\referencecount
   \else \number\lastrefsbegincount-\number\referencecount \fi}\space }
\def\REFNUM#1{\rel@x \gl@bal\advance\referencecount by 1
    \xdef#1{\the\referencecount }}
\def\Refnum#1{\REFNUM #1\refend@ } 
\def\REF#1{\REFNUM #1\R@FWRITE\ignorespaces}
\def\Ref#1{\Refnum #1\REFWRITE }
\def\ref{\Ref\?}
\def\REFS#1{\REFNUM #1\gl@bal\lastrefsbegincount=\referencecount
    \REFWRITE }

\def\r@fitem#1{\par \hangafter=0 \hangindent=\refindent \Textindent{#1}}
\def\refitem#1{\r@fitem{#1.}}
\def\NPrefitem#1{\r@fitem{[#1]}}
\def\NPrefs{\let\refmark=\NPrefmark \let\refitem=\NPrefitem}
\def\REFWRITE{\R@FWRITE\rel@x }
\def\R@FWRITE#1{\ifreferenceopen \else \gl@bal\referenceopentrue
     \immediate\openout\referencewrite=\jobname.refs
     \toks@={\begingroup \refoutspecials \catcode`\^^M=10 }%
     \immediate\write\referencewrite{\the\toks@}\fi
    \immediate\write\referencewrite{\noexpand\refitem %
                                    {\the\referencecount}}%
    \p@rse@ndwrite \referencewrite #1}
\begingroup
 \catcode`\^^M=\active \let^^M=\relax %
 \gdef\p@rse@ndwrite#1#2{\begingroup \catcode`\^^M=12 \newlinechar=`\^^M%
         \chardef\rw@write=#1\sc@nlines#2}%
 \gdef\sc@nlines#1#2{\sc@n@line \g@rbage #2^^M\endsc@n \endgroup #1}%
 \gdef\sc@n@line#1^^M{\expandafter\toks@\expandafter{\deg@rbage #1}%
         \immediate\write\rw@write{\the\toks@}%
         \futurelet\n@xt \sc@ntest }%
\endgroup
\def\sc@ntest{\ifx\n@xt\endsc@n \let\n@xt=\rel@x
       \else \let\n@xt=\sc@n@notherline \fi \n@xt }
\def\sc@n@notherline{\sc@n@line \g@rbage }
\def\deg@rbage#1{}
\let\g@rbage=\relax    \let\endsc@n=\relax
\def\refout{\par\penalty-400\vskip\chapterskip
   \spacecheck\referenceminspace
   \ifreferenceopen \Closeout\referencewrite \referenceopenfalse \fi
   \line{\fourteenrm\hfil REFERENCES\hfil}\vskip\headskip
   \input \jobname.refs
   }
\def\refoutspecials{\sfcode`\.=1000 \interlinepenalty=1000
         \rightskip=\z@ plus 1em minus \z@ }
\def\Closeout#1{\toks0={\par\endgroup}\immediate\write#1{\the\toks0}%
   \immediate\closeout#1}
%
% % % % % % % % % % % % % % % % % % % % % % % % % % % % % % % % % % % %
%%  Next, figure captions and table captions.
%
\newcount\figurecount     \figurecount=0
\newcount\tablecount      \tablecount=0
\newif\iffigureopen       \newwrite\figurewrite
\newif\iftableopen        \newwrite\tablewrite
\def\FIGNUM#1{\rel@x \gl@bal\advance\figurecount by 1
    \xdef#1{\the\figurecount}}
\def\FIGURE#1{\FIGNUM #1\F@GWRITE\ignorespaces }

\def\figitem#1{\r@fitem{#1)}}
\def\FIGWRITE{\F@GWRITE\rel@x }
\def\TABNUM#1{\rel@x \gl@bal\advance\tablecount by 1
    \xdef#1{\the\tablecount}}
\def\TABLE#1{\TABNUM #1\T@BWRITE\ignorespaces }

\def\tabitem#1{\r@fitem{#1:}}
\def\TABWRITE{\T@BWRITE\rel@x }
\def\F@GWRITE#1{\iffigureopen \else \gl@bal\figureopentrue
     \immediate\openout\figurewrite=\jobname.figs
     \toks@={\begingroup \catcode`\^^M=10 }%
     \immediate\write\figurewrite{\the\toks@}\fi
    \immediate\write\figurewrite{\noexpand\figitem %
                                 {\the\figurecount}}%
    \p@rse@ndwrite \figurewrite #1}
\def\T@BWRITE#1{\iftableopen \else \gl@bal\tableopentrue
     \immediate\openout\tablewrite=\jobname.tabs
     \toks@={\begingroup \catcode`\^^M=10 }%
     \immediate\write\tablewrite{\the\toks@}\fi
    \immediate\write\tablewrite{\noexpand\tabitem %
                                 {\the\tablecount}}%
    \p@rse@ndwrite \tablewrite #1}
\def\figout{\par\penalty-400
   \vskip\chapterskip\spacecheck\referenceminspace
   \iffigureopen \Closeout\figurewrite \figureopenfalse \fi
   \line{\fourteenrm\hfil FIGURE CAPTIONS\hfil}\vskip\headskip
   \input \jobname.figs
   }
\def\tabout{\par\penalty-400
   \vskip\chapterskip\spacecheck\referenceminspace
   \iftableopen \Closeout\tablewrite \tableopenfalse \fi
   \line{\fourteenrm\hfil TABLE CAPTIONS\hfil}\vskip\headskip
   \input \jobname.tabs
   }

\newbox\picturebox
\def\p@cht{\ht\picturebox }
\def\p@cwd{\wd\picturebox }
\def\p@cdp{\dp\picturebox }
\newdimen\xshift
\newdimen\yshift
\newdimen\captionwidth
\newskip\captionskip
\captionskip=15pt plus 5pt minus 3pt
\def\fullwidth{\captionwidth=\hsize }
\newtoks\Caption
\newif\ifcaptioned
\newif\ifselfcaptioned
\def\caption{\captionedtrue \Caption }
\newcount\linesabove
\newif\iffileexists
\newtoks\picfilename
\def\fil@#1 {\fileexiststrue \picfilename={#1}}
\def\file#1{\if=#1\let\n@xt=\fil@ \else \def\n@xt{\fil@ #1}\fi \n@xt }
\def\pl@t{\begingroup \pr@tect
    \setbox\picturebox=\hbox{}\fileexistsfalse
    \let\height=\p@cht \let\width=\p@cwd \let\depth=\p@cdp
    \xshift=\z@ \yshift=\z@ \captionwidth=\z@
    \Caption={}\captionedfalse
    \linesabove =0 \picturedefault }
\def\plot{\pl@t \selfcaptionedfalse }
\def\Picture#1{\gl@bal\advance\figurecount by 1
    \xdef#1{\the\figurecount}\pl@t \selfcaptionedtrue }

\def\s@vepicture{\iffileexists \parsefilename \redopicturebox \fi
   \ifdim\captionwidth>\z@ \else \captionwidth=\p@cwd \fi
   \xdef\lastpicture{\iffileexists
        \setbox0=\hbox{\raise\the\yshift \vbox{%
              \moveright\the\xshift\hbox{\picturedefinition}}}%
        \else \setbox0=\hbox{}\fi
         \ht0=\the\p@cht \wd0=\the\p@cwd \dp0=\the\p@cdp
         \vbox{\hsize=\the\captionwidth \line{\hss\box0 \hss }%
              \ifcaptioned \vskip\the\captionskip \noexpand\Tenpoint
                \ifselfcaptioned Figure~\the\figurecount.\enspace \fi
                \the\Caption \fi }}%
    \endgroup }
\let\endpicture=\s@vepicture
\def\savepicture#1{\s@vepicture \global\let#1=\lastpicture }
\def\displaypicture{\fullwidth \s@vepicture $$\lastpicture $${}}
\def\toppicture{\fullwidth \s@vepicture \topinsert
    \lastpicture \medskip \endinsert }
\def\midpicture{\fullwidth \s@vepicture \midinsert
    \lastpicture \endinsert }
%
%  Wraparound macros - a try.
%
\def\leftpicture{\pres@tpicture
    \dimen@i=\hsize \advance\dimen@i by -\dimen@ii
    \setbox\picturebox=\hbox to \hsize {\box0 \hss }%
    \wr@paround }
\def\rightpicture{\pres@tpicture
    \dimen@i=\z@
    \setbox\picturebox=\hbox to \hsize {\hss \box0 }%
    \wr@paround }
\def\pres@tpicture{\gl@bal\linesabove=\linesabove
    \s@vepicture \setbox\picturebox=\vbox{
         \kern \linesabove\baselineskip \kern 0.3\baselineskip
         \lastpicture \kern 0.3\baselineskip }%
    \dimen@=\p@cht \dimen@i=\dimen@
    \advance\dimen@i by \pagetotal
    \par \ifdim\dimen@i>\pagegoal \vfil\break \fi
    \dimen@ii=\hsize
    \advance\dimen@ii by -\parindent \advance\dimen@ii by -\p@cwd
    \setbox0=\vbox to\z@{\kern-\baselineskip \unvbox\picturebox \vss }}
\def\wr@paround{\Caption={}\count255=1
    \loop \ifnum \linesabove >0
         \advance\linesabove by -1 \advance\count255 by 1
         \advance\dimen@ by -\baselineskip
         \expandafter\Caption \expandafter{\the\Caption \z@ \hsize }%
      \repeat
    \loop \ifdim \dimen@ >\z@
         \advance\count255 by 1 \advance\dimen@ by -\baselineskip
         \expandafter\Caption \expandafter{%
             \the\Caption \dimen@i \dimen@ii }%
      \repeat
    \edef\n@xt{\parshape=\the\count255 \the\Caption \z@ \hsize }%
    \par\noindent \n@xt \strut \vadjust{\box\picturebox }}
\let\picturedefault=\relax
\let\parsefilename=\relax
\def\redopicturebox{\let\picturedefinition=\rel@x
   \errhelp=\disabledpictures
   \errmessage{This version of TeX cannot handle pictures.  Sorry.}}
\newhelp\disabledpictures
     {You will get a blank box in place of your picture.}

\def\FRONTPAGE{\ifvoid255\else\vfill\penalty-20000\fi
   \gl@bal\pagenumber=1     \gl@bal\chapternumber=0
   \gl@bal\equanumber=0     \gl@bal\sectionnumber=0
   \gl@bal\referencecount=0 \gl@bal\figurecount=0
   \gl@bal\tablecount=0     \gl@bal\frontpagetrue
   \gl@bal\lastf@@t=0       \gl@bal\footsymbolcount=0}

\def\papers{\papersize\headline=\paperheadline\footline=\paperfootline}
\def\papersize{%xxx \hsize=35pc \vsize=50pc \hoffset=0pc \voffset=1pc
   \advance\hoffset by\HOFFSET \advance\voffset by\VOFFSET
   \pagebottomfiller=0pc
   \skip\footins=\bigskipamount \normalspace }
\papers  %  This is the default
%
% % % % % % % % % % % % % % % % % % % % % % % % % % % % % % % % % % % %
%
\newskip\lettertopskip       \lettertopskip=20pt plus 50pt
\newskip\letterbottomskip    \letterbottomskip=\z@ plus 100pt
\newskip\signatureskip       \signatureskip=40pt plus 3pt
\def\lettersize{\hsize=6.5in \vsize=8.5in \hoffset=0in \voffset=0.5in
   \advance\hoffset by\HOFFSET \advance\voffset by\VOFFSET
   \pagebottomfiller=\letterbottomskip
   \skip\footins=\smallskipamount \multiply\skip\footins by 3
   \singlespace }
\def\MEMO{\lettersize \headline=\letterheadline \footline={\hfil }%
   \let\rule=\memorule \FRONTPAGE \memohead }

\def\memodate{\afterassignment\MEMO \date }
\def\memit@m#1{\smallskip \hangafter=0 \hangindent=1in
    \Textindent{\caps #1}}
\def\subject{\memit@m{Subject:}}
\def\topic{\memit@m{Topic:}}
\def\from{\memit@m{From:}}
%xxx\def\to{\rel@x \ifmmode \rightarrow \else \memit@m{To:}\fi }
\def\memorule{\medskip\hrule height 1pt\bigskip}  % default definitions
\def\memohead{\centerline{\fourteenrm MEMORANDUM}}% see phyzzx.local
\newwrite\labelswrite
\newtoks\rw@toks
\def\letters{\lettersize
   \headline=\letterheadline \footline=\letterfootline
   \immediate\openout\labelswrite=\jobname.lab}

\let\letterhead=\rel@x
\def\addressee#1{\medskip\line{\hskip 0.75\hsize plus\z@ minus 0.25\hsize
                               \the\date \hfil }%
   \vskip \lettertopskip
   \ialign to\hsize{\strut ##\hfil\tabskip 0pt plus \hsize \crcr #1\crcr}
   \writelabel{#1}\medskip \noindent\hskip -\spaceskip \ignorespaces }
\def\rwl@begin#1\cr{\rw@toks={#1\crcr}\rel@x
   \immediate\write\labelswrite{\the\rw@toks}\futurelet\n@xt\rwl@next}
\def\rwl@next{\ifx\n@xt\rwl@end \let\n@xt=\rel@x
      \else \let\n@xt=\rwl@begin \fi \n@xt}
\let\rwl@end=\rel@x
\def\writelabel#1{\immediate\write\labelswrite{\noexpand\labelbegin}
     \rwl@begin #1\cr\rwl@end
     \immediate\write\labelswrite{\noexpand\labelend}}
\newtoks\FromAddress         \FromAddress={}
\newtoks\sendername          \sendername={}
\newbox\FromLabelBox
\newdimen\labelwidth          \labelwidth=6in
\def\makelabels{\afterassignment\Makelabels \sendersname=}
\def\Makelabels{\FRONTPAGE \letterinfo={\hfil } \MakeFromBox
     \immediate\closeout\labelswrite  \input \jobname.lab\vfil\eject}
\let\labelend=\rel@x
\def\labelbegin#1\labelend{\setbox0=\vbox{\ialign{##\hfil\cr #1\crcr}}
     \MakeALabel }
\def\MakeFromBox{\gl@bal\setbox\FromLabelBox=\vbox{\Tenpoint
     \ialign{##\hfil\cr \the\sendername \the\FromAddress \crcr }}}
\def\MakeALabel{\vskip 1pt \hbox{\vrule \vbox{
        \hsize=\labelwidth \hrule\bigskip
        \leftline{\hskip 1\parindent \copy\FromLabelBox}\bigskip
        \centerline{\hfil \box0 } \bigskip \hrule
        }\vrule } \vskip 1pt plus 1fil }
\def\signed#1{\par \nobreak \bigskip \dt@pfalse \begingroup
  \everycr={\noalign{\nobreak
            \ifdt@p\vskip\signatureskip\gl@bal\dt@pfalse\fi }}%
  \tabskip=0.5\hsize plus \z@ minus 0.5\hsize
  \halign to\hsize {\strut ##\hfil\tabskip=\z@ plus 1fil minus \z@\crcr
          \noalign{\gl@bal\dt@ptrue}#1\crcr }%
  \endgroup \bigskip }
\newbox\letterb@x
\def\lettertext{\par \vskip\parskip \unvcopy\letterb@x \par }
\def\multiletter{\setbox\letterb@x=\vbox\bgroup
      \everypar{\vrule height 1\baselineskip depth 0pt width 0pt }
      \singlespace \topskip=\baselineskip }
\def\letterend{\par\egroup}

\newskip\frontpageskip
\newtoks\Pubnum   \let\pubnum=\Pubnum
\newtoks\Pubtype  \let\pubtype=\Pubtype
\newif\ifp@bblock  \p@bblocktrue
\def\PH@SR@V{\doubl@true \baselineskip=24.1pt plus 0.2pt minus 0.1pt
             \parskip= 3pt plus 2pt minus 1pt }
\def\PHYSREV{\papers\PhysRevtrue\PH@SR@V}

\def\titlepage{\FRONTPAGE\papers\ifPhysRev\PH@SR@V\fi
   \ifp@bblock\p@bblock \else\hrule height\z@ \rel@x \fi }
\def\nopubblock{\p@bblockfalse}
\def\endpage{\vfil\break}
\frontpageskip=12pt plus .5fil minus 2pt
\Pubtype={}
\Pubnum={}
\def\p@bblock{\begingroup \tabskip=\hsize minus \hsize
   \baselineskip=1.5\ht\strutbox \topspace-2\baselineskip
   \halign to\hsize{\strut ##\hfil\tabskip=0pt\crcr
       \the\Pubnum\crcr\the\date\crcr\the\pubtype\crcr}\endgroup}
\def\title#1{\vskip\frontpageskip \titlestyle{#1} \vskip\headskip }
\def\author#1{\vskip\frontpageskip\titlestyle{\twelvecp #1}\nobreak}

\def\address#1{\par\kern 5pt\titlestyle{\twelvepoint\it #1}}
\def\andaddress{\par\kern 5pt \centerline{\sl and} \address}

\def\abstract{\par\dimen@=\prevdepth \hrule height\z@ \prevdepth=\dimen@
   \vskip\frontpageskip\centerline{\fourteenrm ABSTRACT}\vskip\headskip }

\def\\{\rel@x \ifmmode \backslash \else {\tt\char`\\}\fi }
\def\sequentialequations{\rel@x \if\equanumber<0 \else
  \gl@bal\equanumber=-\equanumber \gl@bal\advance\equanumber by -1 \fi }
\def\nextline{\unskip\nobreak\hfill\break}

\def\journal#1&#2(#3){\begingroup \let\journal=\dummyj@urnal
    \unskip, \sl #1\unskip~\bf\ignorespaces #2\rm
    (\afterassignment\j@ur \count255=#3), \endgroup\ignorespaces }
\def\j@ur{\ifnum\count255<100 \advance\count255 by 1900 \fi
          \number\count255 }
\def\dummyj@urnal{%
    \toks@={Reference foul up: nested \journal macros}%
    \errhelp={Your forgot & or ( ) after the last \journal}%
    \errmessage{\the\toks@ }}

\def\topspace{\hrule height 0pt depth 0pt \vskip}

\def\Buildrel#1\under#2{\mathrel{\mathop{#2}\limits_{#1}}}
\def\becomes#1{\mathchoice{\becomes@\scriptstyle{#1}}
   {\becomes@\scriptstyle{#1}} {\becomes@\scriptscriptstyle{#1}}
   {\becomes@\scriptscriptstyle{#1}}}
\def\becomes@#1#2{\mathrel{\setbox0=\hbox{$\m@th #1{\,#2\,}$}%
        \mathop{\hbox to \wd0 {\rightarrowfill}}\limits_{#2}}}

\let\int=\intop         
\def\lsim{\mathrel{\mathpalette\@versim<}}
\def\gsim{\mathrel{\mathpalette\@versim>}}
\def\@versim#1#2{\vcenter{\offinterlineskip
        \ialign{$\m@th#1\hfil##\hfil$\crcr#2\crcr\sim\crcr } }}
\def\big#1{{\hbox{$\left#1\vbox to 0.85\b@gheight{}\right.\n@space$}}}
\def\Big#1{{\hbox{$\left#1\vbox to 1.15\b@gheight{}\right.\n@space$}}}
\def\bigg#1{{\hbox{$\left#1\vbox to 1.45\b@gheight{}\right.\n@space$}}}
\def\Bigg#1{{\hbox{$\left#1\vbox to 1.75\b@gheight{}\right.\n@space$}}}
\def\){\mskip 2mu\nobreak }

\let\sec@nt=\sec
\def\sec{\rel@x\ifmmode\let\n@xt=\sec@nt\else\let\n@xt\section\fi\n@xt}
\def\obsolete#1{\message{Macro \string #1 is obsolete.}}
\def\firstsec#1{\obsolete\firstsec \section{#1}}
\def\firstsubsec#1{\obsolete\firstsubsec \subsection{#1}}
\def\thispage#1{\obsolete\thispage \gl@bal\pagenumber=#1\frontpagefalse}
\def\thischapter#1{\obsolete\thischapter \gl@bal\chapternumber=#1}
\def\splitout{\obsolete\splitout\rel@x}
\def\prop{\obsolete\prop \propto }
\def\nextequation#1{\obsolete\nextequation \gl@bal\equanumber=#1
   \ifnum\the\equanumber>0 \gl@bal\advance\equanumber by 1 \fi}
\def\BOXITEM{\afterassigment\B@XITEM\setbox0=}
\def\B@XITEM{\par\hangindent\wd0 \noindent\box0 }

\def\phyzzx{PHY\setbox0=\hbox{Z}\copy0 \kern-0.5\wd0 \box0 X}
        
\everyjob{\xdef\today{\monthname~\number\day, \number\year}
        \input myphyx.tex }
\message{ by V.K.}
%
%xxx\input phyzzx.local
\catcode`\@=12 % at signs are no longer letters
%
%\dump

\sequentialequations
\pubnum{EPHOU-94-002}
\date{August 4, 1994}

\titlepage

\title{Magnetic von-Neumann lattice for two-dimensional electrons
in the magnetic field}

\author{K.Ishikawa, N.Maeda, and K.Tadaki}

\address{Department of Physics
Hokkaido University, Sapporo 060, Japan}

\abstract{
One-particle eigenstates and eigenvalues of two-dimensional electrons
in the strong magnetic field with short range impurity and impurities,
cosine potential, boundary potential, and periodic array of
short range potentials
are obtained by magnetic von-Neumann lattice in which Landau level wave
functions have minimum spatial extensions.
We find that there is a dual correspondence
between cosine potential and lattice kinetic term
and that the representation based on the von-Neumann lattice
is quite useful for solving the system's dynamics.
}
%\submit{Physical Review B}
\endpage

\REF\kl{
K.U.Klitzing, G.Dorda, and M.Pepper, Phys.Rev.Lett.{\bf 45},494(1980).
\nextline
S.Kawaji and J.Wakabayashi, Physics in High Magnetic Fields, ed. S.Chikazami
and N.Miura (Springer-Verlag,Berlin,1981).}

\REF\ts{
D.C.Tsui, H.L.St\"omer, and A.C.Gossard, Phys,Rev.Lett.{\bf 48},1559(1982).}

\REF\ao{
H.Aoki and T.Ando, Solid State Commun.{\bf 38},1079(1981).}

\REF\ab{
E.Abrahams, P.W.Anderson, D.C.Licciardello, and T.U.Ramakrishnun.\nextline
Phys.Rev.Lett.{\bf 42},628(1979).}

\REF\im{
N.Imai, K.Ishikawa, T.Matsuyama, and I.Tanaka, Phys.Rev.B{\bf 42},10610(1990).}

\REF\ku{
R.Kubo, S.J.Miyake, and N.Hashitsume,
Solid State Phys.{\bf 17},269(1965);\nextline
T.Ando, Y.Matsumoto, and Y.Uemura, J.Phys.Soc.Jpan.{\bf 39},272(1975)}

\REF\pe{
A.M.Perelomov, Theor.Mat.Fiz.{\bf 6},213(1971);\nextline
V.Bargmann et al., Rep.Math.Phys.{\bf 2},221(1971).}

\REF\is{
K.Ishikawa, Prog.Theor.Phys.Suppl.No.{\bf 107},167(1992)}

\REF\we{
D.Weiss et al., Phys.Rev.Lett.{\bf 66},2790(1991);\nextline
W.Kamg et al., Phys.Rev.Lett.{\bf 71},3850(1993).}

\doublespace

\chapter{Introduction}

Charged particle changes its property drastically
in the strong magnetic field, especially in two dimensions.
Exciting phenomena such as integer quantum Hall effect$^{\kl)}$
and fractional quantum Hall effect$^{\ts)}$ have been found in this system.
The Hall conductance is regarded to take universal values,
either ${e^2\over h}N$ or ${e^2\over h}{q\over p}$ in these phenomena.
Since these values are universal constants,
it is expected that general physical principle is connected
with the quantum Hall effect.

One of origins of quantum Hall effects is the localization
due to random impurities$^{\ao),\ab)}$
which seems not to be understandable from
perturbative calculations.
Electrons which are localized by impurities have energies
at the outside of Landau level energies.
If the Fermi energy is in this energy region,
there is no zero-energy state and diagonal component
of the electric conductance, ${\sigma}_{xx}$, vanishes.
There is no energy dissipation and field theoretical method
can be applied.
Current conservation is one of the universal relations
and is satisfied in arbitrary physical system.
It leads Ward-Takahashi identity in local field theory.
In quantum Hall system, current conservation is satisfied
but locality of the field for each Landau level
is realized only in suitable representation.
Such a local representation that uses von-Neumann lattice
in center variables was used$^{\im)}$ to derive the exact low energy theorem
for quantum Hall effect.

The representation based on von-Neumann lattice
has not only locality but also translational invariance.
Consequently this representation is convenient
for studying electrons in the magnetic field
with a short range impurity and impurities, boundary,
and periodic array of potentials, as well.
We present these studies in this paper.
We review our representation in Section 2
and show that the localization due to random impurities
can be studied based on perturbative calculations
in our representation in Section 3
and that there is a reciprocal correspondence
between the cosine potential and the lattice kinetic term
in Section 4.
One-particle eigenstates in systems with a boundary potential
and periodic array of short range potentials
are obtained in Section 5.
Summary is given in Section 6.

\chapter{Magnetic von-Neumann lattice$^{\im)}$}

One-body Hamiltonian of a planar charged particle
with a strong perpendicular magnetic field without disorder potential
is given by,
$$
\eqalign{
H=&{({\vec p}+e{\vec A})^2\over 2m},\cr
&\partial_1 A_2-\partial_2 A_1=B.\cr
}
\eqno\eq
$$
The vector potential of the above Hamiltonian is linear
in spatial coordinates and it is useful to introduce
the sets of coordinates,
relative coordinates $(\xi,\eta)$
and guiding center coordinates $(X,Y)^{\ku)}$,
$$
\eqalign{
\xi&={1\over {eB}}(p_y+eA_y),\cr
\eta&=-{1\over {eB}}(p_x+eA_x),\cr
X&=x-\xi,\cr
Y&=y-\eta,\cr
[\xi,\eta]=&-[X,Y]=-{i\hbar\over eB},\cr
[\xi,X]=&[\xi,Y]=[\eta,X]=[\eta,Y]=0.\cr
}
\eqno\eq
$$
One-body Hamiltonian is written with these variables as
$$
H={e^2B^2\over 2m}({\xi}^2+{\eta}^2).
\eqno\eq
$$
Thus we see that the system is equivalent to a harmonic oscillator
and that $X$ and $Y$ are the  constants of the motion.

Many-body problems are studied by a quantized field and the Hamiltonian,
$$
H=\int d\vec x{\Psi}^{\dag}(x){(\vec p+e\vec A)^2\over 2m}\Psi(x).
\eqno\eq
$$
We describe the field theory with the variables of Eq.$(2.2)$.
A set of functions,
$$
\eqalign{
f_l(\xi,\eta)&\otimes\vert R_{m n}\rangle,\cr
f_l(\xi,\eta)&\otimes\widetilde {\langle R_{m n}\vert},\cr
}
\eqno\eq
$$
where $f_l(\xi,\eta)$ is the eigenstate of the harmonic oscillator
and $\vert R_{m n}\rangle$ is the coherent state in $(X,Y)$ variables
are used as basis. $\widetilde{\langle R_{m,n}\vert}$ is a dual basis defined
later.
They satisfy,
$$
\eqalign{
H_0 f_l(\xi,\eta)&={eB\hbar\over m}(l+{1\over 2})f_l(\xi,\eta),\cr
(X+iY)\vert R_{m,n}\rangle&=a(m+in)\vert R_{m,n}\rangle,\cr
{\vec R}_{m n}&=a(m,n),\cr
&m,n\,:\;{\rm integer},\cr
a&=\sqrt{2\pi\hbar\over eB}.\cr
}
\eqno\eq
$$
The coherent states are the minimum wave packets
that have minimum extensions allowed from commutation relations,
Eq.$(2.2)$ and satisfy
$$
\eqalign{
&\Delta X^2\Delta Y^2={{\hbar}^2\over 4e^2B^2},\cr
&\Biggl\{
\eqalign{
\Delta X^2&=\langle(X-\langle X\rangle)^2\rangle\cr
\Delta Y^2&=\langle(Y-\langle Y\rangle)^2\rangle.\cr
}\cr
}
\eqno\eq
$$
A set of the coherent states becomes complete
if the lattice spacing is less than or equal to that of Eq.$(2.6)^{\pe)}$.

Coherent states are not orthogonal despite of their completeness
but satisfy
$$
\eqalign{
&\langle R_{m_1,n_1}\vert R_{m_2,n_2}\rangle=
e^{i\pi[(m_1-m_2+1)(n_1-n_2+1)-1]}e^{-{\pi\over 2}
[(m_1-m_2)^2+(n_1-n_2)^2]},\cr
&\sum_{m_1,n_1}\langle R_{m_1,n_1}\vert R_{m_2,n_2}\rangle=
\sum_{m_2,n_2}\langle R_{m_1,n_1}\vert R_{m_2,n_2}\rangle=0,\cr
}
\eqno\eq
$$
with the following phase convention of the coherent states:
$$
\eqalign{
\vert &R_{m,n}\rangle
=(-1)^{mn+m+n}e^{A^{\dag}\sqrt{\pi}(m+in)-A\sqrt{\pi}(m-in)}\vert 0\rangle,
\cr
&A\vert 0\rangle=0,\cr
&A=\sqrt{eB\over 2\hbar}(X+iY).\cr
}
\eqno\eq
$$
It is convenient to use dual basis that is orthogonal
to the above coherent states
$$
\eqalign{
&\widetilde{\langle R_{m_1,n_1}\vert}R_{m_2,n_2}\rangle=
\delta_{m_1,m_2}\delta_{n_1,n_2}-{1\over N},\cr
&\widetilde{\langle R_{m_1,n_1}\vert}=
\sum_{m_2,n_2} G(m_1,n_1;m_2,n_2)\langle R_{m_2,n_2}\vert,\cr
}
\eqno\eq
$$
$$
\sum_{m',n'} G(m_1,n_1;m',n')\langle R_{m',n'}\vert R_{m_2,n_2}\rangle=
\delta_{m_1,m_2}\delta_{n_1,n_2}-{1\over N},
\eqno\eq
$$
$$
N=\sum_{m_1,n_1}.
\qquad\qquad
$$
The last term in Eqs.$(2.10)$ and $(2.11)$ is due to the constraint
of Eq.$(2.8)$.
The Green's function $G(m_1,n_1;m_2,n_2)$ is obtained
by Fourier transformation,
$$
\eqalign{
G(m_1,n_1;m_2,n_2)&=
\Bigl({1\over 2\pi}\Bigr)^2\int\nolimits_{-\pi}^{\pi}
d\vec P e^{iP_x(m_1-m_2)+iP_y(n_1-n_2)}G(P_x,P_y),\cr
\langle R_{m_1,n_1}\vert R_{m_2,n_2}\rangle&=
\Bigl({1\over 2\pi}\Bigr)^2\int\nolimits_{-\pi}^{\pi}
d\vec P e^{iP_x(m_1-m_2)+iP_y(n_1-n_2)}\alpha(P_x,P_y),\cr
}
\eqno\eq
$$
as
$$
G(P_x,P_y)=
\Biggl\{
\eqalign{
&{1\over \alpha(P_x,P_y)}\qquad\vec P\not=0\cr
&0\qquad\qquad\qquad\vec P=0.\cr}
\eqno\eq
$$
$G(m_1,n_1;m_2,n_2)$ thus calculated is shown in {\bf Fig.1}.
Obviously, it decreases fast with $|m_1-m_2|$ or $|n_1-n_2|$.
Hence the base functions and the dual base functions are
localized functions around center coordinates $R_{m,n}$.
This is an important property of the present base functions
and makes us possible to find spatial properties of the wave functions
in the presence of dis-order potentials
such as short range impurity potentials, boundary potential, and
periodic potentials easily.
Furthermore, since the electric charge of the base function is distributed in
a finite spatial region,
the multi-pole expansion of the charge operator$^{\is)}$
can be defined in the present representation.
The exact low energy theorem$^{\im)}$
was derived based on Ward-Takahashi identity
and the multi-pole expansion.

We expand the electron field $\Psi$ and its conjugate $\Psi^{\dag}$
with these base functions as
$$
\eqalign{
\Psi&=
\sum_{l,m,n}a_l(m,n)
f_l(\xi,\eta)\otimes\vert R_{m n}\rangle,\cr
\Psi^{\dag}&=
\sum_{l,m,n}b_l(m,n)
f_l(\xi,\eta)\otimes\widetilde {\langle R_{m n}\vert},\cr
}
\eqno\eq
$$
and regard $a_l(m,n)$ and $b_l(m,n)$ as quantized operators
\footnote*{
We understand that the operators satisfy,
$\sum_{m,n}b_l(m,n)=\sum_{m,n}a_l(m,n)=0$.
Namely, zero momentum state is not included.
}defined on lattice sites.
The free Hamiltonian, Eq.$(2.4)$, becomes a diagonal form
$$
H=\sum_{l,m,n}E_lb_l(m,n)a_l(m,n).
\eqno\eq
$$
It should be noted that there is no hopping term
in the above Hamiltonian and hence short range potential
is treated locally in this representation.
A potential term is transformed into,
$$
\eqalign{
\int dx\Psi^{\dag}(x)\Psi(x)V(x)&=
\sum b_{l_1}(m_1,n_1)a_{l_2}(m_2,n_2)V_{l_1,l_2}(m_1,n_1;m_2,n_2),\cr
V_{l_1,l_2}(m_1,n_1;m_2,n_2)&=
\int {d^2k\over (2\pi)^2}(f_{l_1}e^{i\vec k\vec {\xi}}f_{l_2})
\widetilde{\langle R_{m_1,n_1}}|e^{i\vec k\vec X}|R_{m_2,n_2}\rangle
V(\vec k),\cr
V(\vec x)=
\int&{d^2k\over (2\pi)^2}e^{i\vec k\vec x}V(\vec k).
\qquad
(\;V(\vec k)=
\int d^2x\>e^{-i\vec k\vec x}V(\vec x)\;)\cr
}
\eqno\eq
$$
$V_{l_1,l_2}(m_1,n_1;m_2,n_2)$ is non-diagonal in spatial directions and
in Landau level indices as well, generally.
If the potential is local in $\vec x$ variable around one position,
then the potential in new variables has the same property,
since the base functions have finite extensions in both directions.
Localization problem due to short range impurities
can be treated based on perturbative expansions.

\chapter{Short range impurity and dilute short range impurities}

Two-dimensional electrons in the magnetic field which interact
with an external potential $V(x)$ are described by
the following Hamiltonian:
$$
\eqalign{
\int d\vec x&\Psi^{\dag}(x)
\left\{{(\vec p+e\vec A)^2\over 2m}+V(x)\right\}
\Psi(x)\cr
&=
\sum_{l_1,l_2}b_{l_1}(m_1,n_1)
\left[E_{l_1}\delta_{l_1,l_2}\delta_{m_1,m_2}\delta_{n_1,n_2}+
V_{l_1,l_2}(m_1,n_1;m_2,n_2)\right]
a_{l_2}(m_2,n_2).\cr
}
\eqno\eq
$$
One-particle properties are found from the eigenvalue equation,
$$
\eqalign{
\sum
[E_{l_1}\delta_{l_1,l_2}\delta_{m_1,m_2}\delta_{n_1,n_2}+
V_{l_1,l_2}(m_1,n_1;m_2,n_2)]&
u_{l_2}^{(\alpha)}(m_2,n_2),\cr
&=E^{(\alpha)}u_{l_1}^{(\alpha)}(m_1,n_1)\cr
\sum v_{l_1}^{(\beta)}(m_1,n_1)
[E_{l_1}\delta_{l_1,l_2}\delta_{m_1,m_2}\delta_{n_1,n_2}+
V_{l_1,l_2}(m_1&,n_1;m_2,n_2)]\cr
&=E^{(\beta)}v_{l_2}^{(\beta)}(m_2,n_2).\cr
}
\eqno\eq
$$
For a short range potential, eigenfunctions $u_{l}^{(\alpha)}(m,n)$
and $v_{l}^{(\alpha)}(m,n)$ decrease
as the distance between the coordinates and
the impurity position increases if the eigenvalue of
the energy $E^{(\alpha)}$ is different from the Landau level energy
$E_l$ from Eq.$(3.2)$ regardless of the sign of the potential.
This is a characteristic feature of the system with the magnetic field
and is shown in {\bf Fig.2}.

We solve the eigenvalue problem of one short range impurity first.
A short range impurity
$$
V(\vec x)=g\delta(\vec x)
\eqno\eq
$$
is transformed to
$$
V_{l_1,l_2}(m_1,n_1;m_2,n_2)
=
g\int {d^2k\over (2\pi)^2}(f_{l_1}e^{i\vec k\vec {\xi}}f_{l_2})
\widetilde{\langle R_{m_1,n_1}}|e^{i\vec k\vec X}|R_{m_2,n_2}\rangle.
\eqno\eq
$$
Eigenfunctions and eigenvalues thus obtained are given in
{\bf Figs.3,4a-4d}.
The eigenfunctions whose eigenvalues are different from Landau level energy
$E_l$ are given in {\bf Figs.4a-4d}
and are obviously localized as is expected from
the equations Eq.$(3.2)$.
The other eigenfunctions have the energy $E_l$ and are
degenerate in energy.
Hence any linear combinations of these eigenfunctions are
also eigenfunctions.
Consequently, the wave functions should be regarded as extended if $E=E_l$
and are localized if $E\not=E_l$,
as is shown in {\bf Fig.2}.

Next we study many but dilute short range impurities
based on perturbative expansions.
We assume that they are distributed randomly
and that their average nearest neighbor distance
is much larger than magnetic distance.
One-body Hamiltonian in magnetic lattice representation is,
$$
\eqalign{
H&=
H_0+\sum_jV_{l_1,l_2}^{(j)}(m_1,n_1;m_2,n_2;\vec {x_j}),\cr
H_0&=
\sum E_{l_1}\delta_{l_1,l_2}\delta_{m_1,m_2}\delta_{n_1,n_2},\cr
V_{l_1,l_2}^{(j)}&(m_1,n_1;m_2,n_2;\vec {x_j})=
g_j\int {d^2k\over (2\pi)^2}
(f_{l_1}(\xi,\eta)e^{i\vec k\vec {\xi}}f_{l_2}(\xi,\eta))
\widetilde{\langle m_1,n_1}|e^{i\vec k\vec X}|m_2,n_2\rangle
e^{-i\vec k\vec {x_j}},\cr
}
\eqno\eq
$$
$$
|\vec {x_j}-\vec {x_{j'}}|\gg a.
\eqno\eq
$$
Since distances between impurities are much larger than
the magnetic distance, each term
in the second part of the Hamiltonian approximately commutes.
Perturbative treatment of many impurity system
is thus possible.

{\bf Theorem 1.}

Characteristic properties of localized eigenstate with discrete eigenvalue,
$E_0^{(\alpha)}$, and with localized wave function around one impurity,
$$
\eqalign{
H_1u^{(\alpha)}&=E_0^{(\alpha)}u^{(\alpha)},\cr
H_1&=H_0+V_{l_1,l_2}^{(i)}(m_1,n_1;m_2,n_2;\vec {x_i}),\cr
}
\eqno\eq
$$
are unchanged by the perturbative terms due to other impurities,
$$
H_2=\sum_{j\not=i}V^{(j)}(m_1,n_1;m_2,n_2;\vec {x_j}),
\eqno\eq
$$
if the energy $E_0^{(\alpha)}$ satisfies
$$
|E_0^{(\alpha)}-E_l|>\delta,
\eqno\eq
$$
where $\delta$ is a small constant determined from impurity distributions.

{\bf Proof}

Perturbative treatment of $H_2$ for the eigenstates and eigenvalues
of total Hamiltonian leads the following expansions:
$$
\eqalign{
E^{(\alpha)}&=
E_0^{(\alpha)}+E_1^{(\alpha)}+E_2^{(\alpha)}+\cdots,\cr
E_1^{(\alpha)}&=<\alpha|H_2|\alpha>,\cr
E_2^{(\alpha)}&=\sum_{\beta\not=\alpha}
{|<\alpha|H_2|\beta>|^2\over {E_0^{(\alpha)}-E_0^{(\beta)}}},
\qquad E_3^{(\alpha)}=\cdots,\cr
}
\eqno\eq
$$
$$
\eqalign{
u^{(\alpha)}&=
u_0^{(\alpha)}+u_1^{(\alpha)}+u_2^{(\alpha)}+\cdots,\cr
u_1^{(\alpha)}&=
\sum_{\beta\not=\alpha}
{<\alpha|H_2|\beta>\over {E_0^{(\alpha)}-E_0^{(\beta)}}}
u_0^{(\beta)},\cr
u_2^{(\alpha)}&=
\sum_{\beta\not=\alpha}
\sum_{\gamma\not=\alpha}
{<\gamma|H_2|\beta><\beta|H_2|\alpha>
\over {(E_0^{(\alpha)}-E_0^{(\beta)})(E_0^{(\alpha)}-E_0^{(\gamma)})}}
u_0^{(\gamma)}-
\sum_{\gamma\not=\alpha}
{<\alpha|H_2|\alpha><\gamma|H_2|\gamma>
\over {(E_0^{(\alpha)}-E_0^{(\gamma)})^2}}
u_0^{(\gamma)}\cr
&\qquad-{1\over 2}
\sum_{\gamma\not=\alpha}
{|<\alpha|H_2|\gamma>|^2\over {(E_0^{(\alpha)}-E_0^{(\gamma)})^2}}
u_0^{(\gamma)}.\cr
}
\eqno\eq
$$
If the state $\alpha$ is the localized state,
the energy denominator is finite
and the matrix element $<\beta|H_2|\alpha>$ is of order
$e^{-{\pi\over 2}({|\vec{x_i}-\vec{x_j}|\over a})^2}$, which is very small
from the condition
that the distance $|\vec {x_i}-\vec {x_j}|$ is much larger
than the magnetic distance.
Hence the perturbative series converges well.
Consequently, the energy eigenvalue $E^{(\alpha)}$ is about the same
as $E_0^{(\alpha)}$.
In {\bf Fig.5}, we show first order correction $E_1^{(\alpha)}$
for randomly distributed potential $V_2$.
First order correction to the wave function, $u_1^{(\alpha)}$,
is also very small
if the state $|\beta>$ is also the localized state
around $i$-th impurity
and is small if the state $|\beta>$ is the degenerate state
with energy $E_l$ and its wave function is finite
around other impurities.
Thus the correction $u_1^{(\alpha)}$ to the wave function
is negligibly small generally.
In the exceptional case
where $|E_0^{(\alpha)}-E_l|$ is infinitesimally small,
the coefficient ${<\alpha|H_2|\beta>\over {E_0^{(\alpha)}-E_0^{(\beta)}}}$
could be order $1$,
and the correction to the wave function becomes larger.
This occurs if
$$
|E_0^{(\alpha)}-E_l|
\mathop{<}_{\approx}
ge^{-{\pi\over 2}({L\over a})^2},
\eqno\eq
$$
where $L$ is the impurity distance and $g$ is the impurity strength,
is satisfied.

{\bf Theorem 2.}

Degenerate states with the energy $E_l$ in the presence of $i$-th impurity
get large corrections from another $j$-th impurity
if wave functions are finite around that impurity.
Degenerate perturbations is applied in this case.

{\bf Proof}

In the ordinary perturbative  formula, the energy denominator vanishes
if $E_0^{(\alpha)}=E_l$ and $E_0^{(\beta)}=E_l$
and the direct diagonalization in the space of energy $E_l$
and of having finite matrix element $<\beta|H_2|\gamma>$
should be made, instead of ordinary perturbative expansions.
Then we have,
$$
[H_0+g_iV_{l_1,l_2}^{(i)}(m_1,n_1;m_2,n_2;\vec {x_i})]
u_{l_2}^{(\alpha)}(m_2,n_2)=
E_lu_{l_1}^{(\alpha)}(m_1,n_1),
\eqno\eq
$$
and
$$
g_jV_{l_1,l_2}^{(j)}(m_1,n_1;m_2,n_2;\vec {x_j})
u_{l_2}^{(\alpha)}=
\delta E^{\alpha}
u_{l_1}^{(\alpha)}.
\eqno\eq
$$
Combining both equations, eigenstate of $V^{(j)}$ in degenerate space satisfy,
$$
\eqalign{
[H_0+
g_iV_{l_1,l_2}^{(i)}(m_1,n_1;m_2,n_2;\vec {x_i})+
g_j&V_{l_1,l_2}^{(j)}(m_1,n_1;m_2,n_2;\vec {x_j})]
u_{l_2}^{(\beta)}(m_2,n_2)\cr
&=(E_l+\delta E^{\alpha})
u_{l_1}^{(\beta)}(m_1,n_1).\cr
}
\eqno\eq
$$

{\bf Theorem 3.}

Localized eigenstates from the degenerate perturbations in theorem 2
have their wave functions around $j$-th impurity.
They can be computed, alternatively, by solving localized eigenstates
of $H_0+g_jV^{(j)}$ first and by applying perturbative expansion
with respect to $\sum_{i\not=j}g_iV^{(i)}$ second.

{\bf Proof}

Let $u^{(\beta)}$ be eigenstate of total Hamiltonian
and $u_0^{(\alpha)}$ be eigenstate of $g_iV^{(i)}$
within the space of degenerate energy,
$$
\eqalign{
[H_0+g_jV^{(j)}+&\sum_ig_iV^{(i)}]u^{(\beta)}=E^{(\beta)}u^{(\beta)},\cr
g_iV^{(i)}u_0^{(\alpha)}&=\delta E^{\alpha}u_0^{(\alpha)}+\delta u,\cr
(H_0+g_jV^{(j)})&u_0=E_lu_0,\qquad\qquad(u_0,\delta u)=0.\cr
}
\eqno\eq
$$
{}From the second and the third equations, $u_0$ satisfies approximately
the eigenvalue equation of total Hamiltonian.
We can write
$$
u^{(\beta)}=u_0+{\delta}'u.
\eqno\eq
$$
The distance between $i$-th impurity and $j$-th impurity is large,
hence $u_0$ is further given approximately
by $l$-th Landau level wave functions,
$$
\eqalign{
u_0&=u_l+{\delta}'' u,\cr
H_0&u_l=E_lu_l.\cr
}
\eqno\eq
$$
Consequently, we have the expression,
$$
\eqalign{
u^{(\beta)}&=u_l+{\delta}'u+{\delta}''u,\cr
E^{(\beta)}&=E_l+\delta E^{(\beta)}.\cr
}
\eqno\eq
$$
Within the $l$-th Landau level space, $u_l$ is the eigenstate
of $H_0+g_jV^{(j)}$,
$$
[H_0+g_jV^{(j)}]u_l=(E_l+\delta E)u_l+\delta u_{l'}\qquad(l'\not=l).
\eqno\eq
$$
Hence the localized eigenfunctions around $j$-th impurity
can be obtained by perturbative expansions.
Eigenvalue $E^{(\beta)}$ thus obtained is generally different
from the discrete eigenvalue $E^{(\alpha)}$ of the eigenstate
around $i$-th impurity,
since these impurities are random without any correlations.
Two eigenvectors are orthogonal, if their eigenvalues are different.
Hence these eigenvectors are linearly independent.

{\bf Theorem 4.}

In a system of dilute impurities, the localized states around $i$-th impurity
of having isolated non-degenerate energies are obtained
from the localized eigenfunctions of $H_0+V^{(i)}$
and their corrections due to $\sum_{l\not=i}V^{(l)}$.
The expansion converges well and the corrections are small,
from theorem 1, if the energy satisfies
$$
\eqalign{
&|E-E_l|\gg\delta,\cr
&\delta\simeq ge^{-{\pi\over 2}({L\over a})^2}.\cr
}
\eqno\eq
$$

{\bf Proof}

Theorem 2 and 3 give theorem 4.

{\bf Theorem 5.}

Since the localized states around each impurity are treated perturbatively,
the number of the localized states are proportional to the number
of the impurities in the dilute impurity cases.

{\bf Proof}

Theorem 2, 3, and 4 give theorem 5.

Combining theorems 1, 2, 3, 4, and 5,
we obtain the one-particle property in the dilute impurity case
and show it in {\bf Fig.6}.

\chapter{Cosine potential: a duality relation between
 potential term and kinetic term}

In this section we study simple cosine potential,
$$
V(x)=V_0(\cos kx+\cos ky),
\eqno\eq
$$
as a special example of periodic potentials
and we show that the above potential term in original representation
is transformed to the hopping term in magnetic lattice representation.
Thus a kind of duality relation
between the kinetic term and the potential term
is observed in the two-dimensional system
with the perpendicular magnetic field.
We saw that the kinetic term is transformed
to the mass term in Section 2.
Now we see that the potential term
$$
\int d\vec x V(\vec x)\Psi^{\dag}(x)\Psi(x)
\eqno\eq
$$
with the periodic potential, Eq.$(4.1)$,
is transformed to the lattice kinetic term,
$$
\sum_{l,\vec R}K_lb_l(\vec R)a_l(\vec R+\vec \Delta)+
{\rm inter\ Landau\ level\ term},
\eqno\eq
$$
$$
\eqalign{
K_l&=
{1\over 2}{(-1)^l\over{l!}}(\pi\Delta n^2)^l
e^{-{\pi\over 2}\Delta n^2}(-1)^{\Delta n},\cr
\vec {\Delta}&=
(\pm\Delta n,0),\;(0,\pm\Delta n),\cr
\Delta n&=
k\sqrt{\hbar\over{2eB\pi}},\cr
}
\eqno\eq
$$
if the $\Delta n$ defined in the above equation is integer.
The energy eigenvalue within the lowest Landau level space
becomes, thus,
$$
\eqalign{
E(\vec k)=
E_0+2K_0(\cos k_xa+\cos k_ya)&,\cr
-{\pi\over a}\leq k_i\leq{\pi\over a}&.\cr
}
\eqno\eq
$$
The energy has the same form as the potential Eq.$(4.1)$.
The duality relation between the kinetic term and the potential term
is summarized in {\bf Table 1}.

\chapter{Degenerate systems: boundary potential and periodic array
of short range potentials.}

If the impurity system has some symmetry,
there is an additional degeneracy.
Naive perturbative expansion is not valid, then, and Hamiltonian
is diagonalized directly.
We study such systems in this section.

\section{boundary potential}

Potential
$$
V(\vec x)=V_0\theta(x)
\eqno\eq
$$
shows a potential barrier in positive $x$ region with a boundary at $x=0$.
Electron stays in negative $x$ region due to the potential barrier if its
energy is less than $V_0$. The potential is invariant under translation in
y-direction and the momentum in y-direction, $P_y$, is a good quantum number.
The single particle energy becomes $P_y$ dependent. We solve the eigenvalue
equation using the magnetic lattice representation. The potential and the
eigenvectors are given by
$$
\eqalign{
V_{l_1,l_2}(m_1,n_1;m_2,n_2)&=
{a\over 2\pi}\int\nolimits_{-{\pi\over a}}^{\pi\over a}dP_y
e^{iP_y(n_1-n_2)a}V_{l_1,l_2}(m_1;m_2;P_y)
,\cr
u_l(m,n)&=\Bigl({a\over 2\pi}\Bigr)^{1\over 2}
\int\nolimits_{-{\pi\over a}}^{\pi\over a}dP_y
e^{iP_yn a}u_l(m,P_y).
\cr
}
\eqno\eq
$$
We substitute the above formula to the eigenvalue equation and we have
$$
\sum_{l_2,m_2}
\{E_{l_1}\delta_{l_1,l_2}\delta_{m_1,m_2}+
V_{l_1,l_2}(m_1;m_2;P_y)\}
u_{l_2}(m_2,P_y)
=
Eu_{l_1}(m_1,P_y).
\eqno\eq
$$
The above equation is solved numerically. The energy and the eigenvector thus
obtained are shown in {\bf Fig.7a-7c}. The solutions are classified into three
types. First type solutions have the energy of Landau levels, and have
the wave functions at $x\ll 0$. Solution of the second type have the energy
$E_l+V_0$ and have the wave functions at $x\gg 0$. Third type solutions have
the energy in the intermediate range, between $E_l$ and $E_l+V_0$ and have the
wave functions at around boundary, $x=0$. The last solutions are called edge
states and play some special roles in electric conduction, especially when
$V_0$ is very large.
The single-particle
energy is periodic in $P_y$, since the electron system is defined on lattice
sites. Consequently, there are even number of solutions in the equation,
$$
\eqalign{
&E(P_F)=E_F,\cr
&E_l<E_F<E_l+V_0.\cr
}
\eqno\eq
$$

One of solution has a steep slope and hence there are small number of states
in this region. Their effect becomes negligible in $V_0\rightarrow
\infty$ limit.

The edge states carry electric current. In the present representation, the
velocity operators are expressed by relative coordinates $\xi$ and $\eta$
which are independent from the motion in center variables. Hence Landau level
mixing is necessary for the states to have finite current. The current
carried by each state is shown in {\bf Fig.8}.

The solutions are extended in y-direction but are localized  in x-direction
and have zero energy. The physical effects of these one-dimensional zero
energy states in electric conductance shall be discussed in a forthcoming
paper.

\section{periodic array of short range potentials}

We study the electron system of having periodic array of short range
potentials in this section. The potential,
$$
V(\vec x)=\sum_jg\delta(\vec x-\vec{x_j}),\qquad
\vec{x_j}=b(j_1,j_2),
\eqno\eq
$$
satisfies,
$$
\eqalign{
&V(\vec x+\vec b)=V(\vec x),\cr
&\vec b=b(m,n)\qquad m,n\;{\rm integer}.\cr
}
\eqno\eq
$$
The magnetic translation operator,
$$
T(\vec b)=
e^{i(\vec p+e\vec A)\cdot\vec b+i\vec b\cdot e\vec {B_z}\times\vec x},
\eqno\eq
$$
then commutes with the Hamiltonian, and satisfies, furthermore,
$$
T(\vec {b_1})T(\vec {b_2})=
\exp\Bigl[ie\vec {B_z}\cdot(\vec {b_1}\times\vec {b_2})\Bigr]
T(\vec {b_2})T(\vec {b_1}).
\eqno\eq
$$

In the present von-Neumann lattice, the electorns are defined on lattce sites
with lattice spacing $a=\sqrt{2\pi\hbar\over eB}$. Since the phase factor in
Eq.$(5.8)$ is given by
$$
2\pi(m_1n_2-m_2n_1),
\eqno\eq
$$
the magnetic translation
group becomes Abelian group in this lattice space
and the translation operators commute.

The electron system's property depends drastically upon the ratio between the
period of the potential and the spacing $a$. The system is translational
invariant if the ratio is a rational number with two integers,
$$
{b\over a}={q\over p},
\eqno\eq
$$
but is non-invariant if the ratio is an irrational number. We compute the
single-particle energies and show them in
{\bf Fig.9a,9b}.
The effect of the potentials becomes large if the ratio is equal to one. The
spacing of the periodic potentials, then is equal to
643[$\AA$] if $B=1\,${\bf T}
or 203[$\AA$]
if $B=10\,${\bf T}. These values are much larger than the original
lattice spacing of the matter, and these potentials should be made
artificially.
They are achieved actually and interesting observations are made$^{\we)}$.

\chapter{Summary and outlook}

We have shown that the two-dimensional electron systems with the
perpendicular magnetic field are studied easily using von-Neumann
magnetic lattice
representation in center coordinates. Both of locality and translational
invariance are explicitly realized and hence multi-pole expansion of charge
operator is possible. Since the spatial properties of
potentials are reflected directly to the equations,
electron systems with disorder potentials are studied
easily with the present representation. Localization problem in
short range impurities was solved  based on perturbative expansions in the
present representation. This was possible because the base functions are
localized functions and localized eigenfunctions around short range
impurities are expressed in straightforward manner.

Electrons in the perpendicular magnetic field are expressed on the lattice in
our representation. It was shown that the kinetic term is transformed to the
mass term and conversely the cosine potential term is transformed to the
lattice kinetic term.
Eigenstates in the system with the boundary and
the periodic array of short range
impurities were obtained in Section 5. Edge states, which are extended along
the edge but are localized in the perpendicular direction, are found and their
energies are computed as a periodic function of the momentum along the
boundary in the former system. The latter system changes its property
drastically when the period of the potentials is changed. They will be
observed experimentally.

{\bf Acknowledgement}

The present work is partially supported
by the special Grant-in-Aid for Promotion of Education and Science in Hokkaido
University Provided by the Ministry of Education, Science and Culture,
a Grant-in-Aid for General Scientific Reseach (03640256),
and the Grant-in-Aid for Scientific Research on Priority Area (043231101),
the Ministry of Education, Science and Culture, Japan. One of the authers
(K,I) thanks Professers A.Zee and E.D'Hoker for usefull discussions.

\endpage

\centerline{FIGURE CAPTIONS}
\leftskip 0.6 truecm
\medskip
\item{{\bf Fig.1:}}{
Green's function $G(m,n;0,0)$, which decreases very fast with $|m|$ and $|n|$,
is shown.}

\item{{\bf Fig.2:}}{
Eigenfunctions of Eq.$(3.2)$ should be localized functions if their energies
are different from those of pure Landau levels.}

\item{{\bf Fig.3:}}{
Energy eigenvalues in the system of one short range impurity are
shown. Discrete eigenvalues are changed almost linearly with coupling
strength. $g_r$ is given by ${2eB\over \hbar}{g\over(2\pi)^2}
({eB\hbar\over m})^{-1}$.}

\item{{\bf Fig.4:}}{
Eigenfunctions with non-degenerate eigenvalues of a short range impurity
system are shown. Lower semi-circles stand for the magnitudes of the $l=0$
components at each lattice sites and upper semi-circles stand for the
magnitudes of the $l=1$ components at each lattice sites. {\bf Fig.4a} and
{\bf 4b} correspond to energy eigenvalue 0.8318 and 2.412
for $g_r=0.2$ respectively. {\bf Fig.4c} and {\bf 4d} correspond to
0.6116 and 1.637 for $g_r=0.04$.
It is obvious that these functions are
localized functions.}

\item{{\bf Fig.5:}}{
Discrete energies are slightly modified by lowest order corrections due to
a nearby impurity. Discrete energies arround one short range impurity,
$E^{(\alpha)}_0$, and their lowest order corrections due to a nearly
impurity, $E^{(\alpha)}_1$, are shown.
Impurity strength is 0.2 in {\bf Fig.5a} and 0.04 in
{\bf Fig.5b}. Distance between impurities is assumed to be one lattice
spacing, $a$.}

\item{{\bf Fig.6:}}{
Energy levels of dilute short range impurity system are shown. Eigenfunctions
are localized if their energy satisfies $|E-E_l|>\delta$, where $\delta$ is a
constant determined by impurities.}

\item{{\bf Fig.7:}}{
Eigenfunctions of the system with a boundary are shown. Edge state energy
becomes large at some momentum along the edge.The potential barrier
is $5$ times as large as that of Landau level energy spacing in {\bf Fig.7a}
and $10$ times as large as that of Landau levels in {\bf Fig.7b}. Wave
functions for $P_y=0$ and $V_0=0.8\;[{eB\hbar\over m}]$
 are shown in {\bf Fig.7c}.}

\item{{\bf Fig.8:}}{
The currents carried by the edge states are shown for three different
potential heights($V_0=0.8,\;5,\;10\;[{eB\hbar\over m}]$).
They are negative definite as is expected from a naive
consideration.}

\item{{\bf Fig.9:}}{
The energy distributions in the periodic array of potentials vs. the ratio
between the potential period, $b$, and magnetic distance, $a$, are shown. In
{\bf Fig.9b}, the same figure of the region, ${b\over a}<1.5$, is shown.}

\endpage

\refout

\end
%%%%%%%%%%%%%%%%%%%%%%%%%%%%%%%%%%%%%%%%%%%%%%%%%%%%%%%%
% LaTeX file for Table                                 %
%%%%%%%%%%%%%%%%%%%%%%%%%%%%%%%%%%%%%%%%%%%%%%%%%%%%%%%%
\documentstyle[12pt]{article}

\begin{document}
\pagestyle{empty}
\baselineskip=20pt

Table 1.
Correspondence between the system of zero-magnetic field

and the system of strong magnetic field.

\

\begin{tabular}{|c|c|c|}
\hline
$\quad B=0\quad$&$\quad
\int d\vec x\psi^\dagger(x){(\vec p+e\vec A)^2\over 2m}\psi(x)\quad$
&$\quad\int d\vec x V(x)\psi^\dagger(x)\psi(x)\quad$\\
     & kinetic term & potential term \\ \hline
$B\neq0$ &$\sum E_l b_l a_l(\vec R)$&$\sum Kb_l(\vec R)a_l(
\vec R+\vec\Delta)$\\
         & mass term & kinetic term \\ \hline
\end{tabular}
\end{document}